\documentclass[12pt]{article}

\usepackage{times}

\usepackage{graphicx}

\newenvironment{sciabstract}{%
\begin{quote} \bf}
{\end{quote}}

\newcounter{lastnote}

\title{Spin Wave Lifetimes Throughout the Brillouin Zone}

\author
{S.P. Bayrakci,$^{1\ast}$ T. Keller,$^{1,2}$ K. Habicht,$^{3}$ and B. Keimer$^{1}$\\
\\
\normalsize{$^{1}$Max-Planck-Institut f\"ur Festk\"orperforschung, Heisenbergstr. 1, 70569 Stuttgart, Germany}\\
\normalsize{$^2$ ZWE FRMII, Technische Universit\"at M\"unchen,
Lichtenbergstr. 1, 85748 Garching, Germany}\\
\normalsize{$^{3}$Hahn-Meitner-Institut, Glienickerstr. 100, 14109 Berlin, Germany}\\
\\
\normalsize{$^\ast$To whom correspondence should be addressed;
E-mail:  bayrakci@fkf.mpg.de.} }

\date{}

\begin{document}


\baselineskip24pt


\maketitle


\begin{sciabstract}
We use a neutron spin-echo method with $\mu$eV resolution to
determine the lifetimes of spin waves in the prototypical
antiferromagnet MnF$_2$ over the entire Brillouin zone. A theory
based on the interaction of magnons with longitudinal spin
fluctuations provides an excellent, parameter-free description of
the data, except at the lowest momenta and temperatures. This is
surprising, given the prominence of alternative theories based on
magnon-magnon interactions in the literature. The results and
technique open up a new avenue for the investigation of
fundamental concepts in magnetism.  The technique also allows
measurement of the lifetimes of other elementary excitations (such
as lattice vibrations) throughout the Brillouin zone.
\end{sciabstract}
\newpage

%A large body of theoretical work is available, but could not be
%tested experimentally for lack of a momentum-resolved method with
%sufficient energy resolution.

The concept of elementary excitations is one of the basic pillars
of the theory of solids. In the low-temperature, long-wavelength
limit, such excitations do not interact and have an infinite
lifetime. For nonzero temperatures and momenta, the lifetimes of
elementary excitations are generally limited by collisions with
other excitations, with important consequences for the macroscopic
properties of solids. For instance, the thermal expansion of
solids can be understood as a consequence of collisions between
lattice vibrations (phonons). Because of their comparatively
simple Hamiltonians, magnetically-ordered states are excellent
testing grounds for theories of elementary excitations and their
interactions.  Despite this, the damping of spin waves (magnons)
in antiferromagnets has remained an open problem for four decades.
Theoretical calculations of magnon lifetimes have been carried out
since the 1960's, with intensive development occurring on several
fronts in the early 1970's. However, these activities ground to a
halt by the mid-1970's due to the lack of appropriate experimental
data, namely, from momentum-resolved measurements with sufficient
energy resolution.  The only low-temperature data available were
taken with $q \cong 0$, in antiferromagnetic resonance (AFMR) and
parallel pumping measurements \cite{kotthaus73,barak80}. Because
of the limited energy resolution, momentum-resolved data from
neutron spectroscopy \cite{schulhof71}, on the other hand, were
confined to the critical regime extremely close to the N\'eel
temperature ($\rm T_N$), where most theories do not apply. Until
recently, no other experimental techniques were available which
permitted high-resolution measurements of excitation lifetimes at
low temperatures over the whole Brillouin zone.  We report on a
new neutron spectroscopy method with $\mu$eV resolution which is
used to measure spin wave (magnon) lifetimes in the prototypical
antiferromagnet MnF$_2$ over the temperature range from 0.04 - 0.6
$\rm T_N$. The results subject long-standing theoretical
predictions to a first experimental test and hold promise as a
novel probe of elementary excitations in quantum magnets.  The
technique is also widely applicable to other elementary
excitations such as phonons and crystal field excitations.

The determination of magnon lifetimes at low temperatures requires
an energy resolution in the $\mu$eV range, about two orders of
magnitude better than that achievable by standard neutron
triple-axis spectroscopy (TAS). We have obtained the requisite
gain in resolution by manipulating the Larmor phase of the neutron
spin with magnetic fields.
%Fig. 1A shows a sketch of the
%experimental setup at the TRISP spectrometer at the FRM-II
%research reactor in Garching, Germany \cite{keller02}.
The TRISP spectrometer (Fig. 1A) weds the capability of TAS of
accessing collective excitations throughout the Brillouin zone to
the extremely high energy resolution of neutron spin-echo
spectroscopy\cite{keller02}.  As in typical spin-polarized
triple-axis spectrometry, the neutrons impinging upon the sample
are polarized, and the polarization of neutrons scattered from the
sample is measured. On TRISP, this is accomplished through the use
of a polarizing neutron guide and a transmission polarizer,
respectively. However, in analogy to neutron spin-echo
spectrometry, the TRISP spectrometer also includes regions of
effectively constant magnetic field which are produced by pairs of
radio-frequency (RF) resonance coils inserted symmetrically 1)
between the monochromator and sample and 2) between the sample and
the analyzer \cite{golub87}. The RF frequencies in the coils are
tuned such that each detected neutron that creates an excitation
lying on the magnon dispersion curve has the same net Larmor phase
after traversing the two spin-echo arms, independent of small
variations in the wave vector of the excitation. The neutron spin
polarization determined at the detector is then a measure of the
linewidth (inverse lifetime) of the magnon. In this way, the
measured linewidth is decoupled (to first order) from the spread
in energy of the neutrons incident on the sample, which is
responsible for the instrumental resolution in TAS. (For a
detailed description of the technique, see the Materials and
Methods section.)

We chose the antiferromagnet MnF$_2$ for the experiment, as its
magnetic ground state and excitations have been investigated
extensively.  The lattice structure and magnetic ordering of
MnF$_2$ are shown in Fig. 1B. MnF$_2$ has the body-centered
tetragonal structure, with a = b = 4.8736 \AA\ and c = 3.2998 \AA.
The Mn$^{2+}$ ions have spin S = 5/2, and the spin in the center
of the unit cell is oriented antiparallel to those at the corners.
The strongest magnetic interaction is between
second-nearest-neighbor Mn$^{2+}$ spins (corner and center spins)
and is antiferromagnetic \cite{okazaki64}.  A weaker,
ferromagnetic interaction exists between nearest-neighbor spins
(along the $c$-axis).  A relatively strong uniaxial anisotropy
which is predominantly the result of dipole-dipole interactions
\cite{keffer52,barak78} causes the spins to align along the
$c$-axis. $\rm T_N$ is 67.6 K. The slope of the magnon dispersion
is required to set the tilt angles of the RF coils and to
determine the non-intrinsic contribution to the data (see
Materials and Methods). During the experiment, the spin-wave
dispersion was therefore measured at each temperature at which
linewidth data was taken; a partial data set is shown in Fig. 1C.

Fig. 2 shows raw polarization data as a function of the spin-echo
time $\tau$.  The spin-echo time is proportional to the frequency
in the RF coils and the distance between the coils, and also
depends on the neutron wavelength. In a neutron spin-echo
experiment, the dependence of the measured polarization on $\tau$
corresponds to the Fourier transform of the scattering function as
a function of energy.  The data in Fig. 2 are described well by an
exponential decay, which indicates that the spectral function
which characterizes the magnon linewidth is a Lorentzian in
energy.  The difference in linewidth (half-width at half maximum,
or HWHM) between the upper two and lower two data sets is in each
case only $\sim\!$ 3 $\mu$eV, but it can be resolved clearly. The
upper pair of data sets represents a difference in $q$ of 0.05
r.l.u. at 15 K \cite{units}. For comparison, the HWHM of the
corresponding TAS scans of the lower two magnons in Fig. 2, taken
with fixed final neutron wave vector $k_{f}$=1.7 \AA$^{-1}$, is
approximately 100 $\mu$eV.

The raw data were then corrected for instrumental and
non-intrinsic effects \cite{materialsandmethods}. Figs. 3 and 4
show the intrinsic magnon linewidth as a function of momentum $q$
and temperature T, respectively. The linewidth generally increases
with increasing $q$ and T, due to the increasing likelihood of
collisions with other excitations. However, Fig. 3 also shows that
the linewidth deviates from this general trend and exhibits peaks
as a function of $q$ close to the center and the boundary of the
antiferromagnetic Brillouin zone. The low-$q$ peak is already
present at 3 K, the lowest temperature covered by this experiment,
and it evolves weakly with increasing temperature. This behavior
is not described by the dominant magnon relaxation mechanisms for
which quantitative predictions are available; possible origins
will be discussed below. In order to facilitate comparison with
these predictions, we have treated the 3 K linewidth data as a
temperature-independent contribution and subtracted it from the
higher-temperature data. The results are shown in the main panels
of Figs. 3 and 4.

The intrinsic relaxation channel for magnons that has received by
far the most attention in the literature is magnon-magnon
scattering. In an ``$n$-magnon" scattering event, a magnon (here,
one excited by an incoming neutron) scatters off $(n/2-1)$
thermally excited magnons, producing $n/2$ scattered magnons which
are in thermal equilibrium with the sample. In the absence of
defects and external magnetic fields, the lowest-order interaction
which limits the magnon lifetime is 4-magnon scattering.
Unfortunately, a comprehensive survey of the literature revealed
very few theoretical predictions appropriate for comparison with
our data, despite the existence of considerable work on 4- and
6-magnon interactions. This is either because the calculations
employed approximations valid only in high magnetic fields (for
the purpose of comparison with AFMR data), or because strict
inequalities that define the range of applicability of the
theoretical results are extremely difficult to satisfy
experimentally. An analytical expression was given by Harris {\it
et al.} \cite{harris71}, who evaluated the contribution to the
linewidth from 4-magnon scattering processes analytically for the
case of single-ion anisotropy with $q = 0$ \cite{harris71note}.
The corresponding result is shown in the bottom trace of Fig. 4.
At low temperatures, the temperature dependence of the data is
considerably weaker than that predicted by this theoretical result
for $q = 0$. The best agreement of the magnitude occurs at 40 K,
where the experimental result is $ \sim 30$\% larger than the
theoretical. Predictions for an anisotropy gap of dipolar origin,
which would be more appropriate for MnF$_2$, are not available.

An additional relaxation channel, in which magnons are scattered
by thermally-excited longitudinal spin fluctuations, was
considered by Stinchcombe and coworkers \cite{cs71b,stinch74}. The
curves in Figs. 3 and 4 are based on this mechanism.  (For $q=0$,
where the contribution of this relaxation is identically zero, we
have shown the prediction of the 4-magnon relaxation model, as
discussed above.)   For the larger-$q$ data, the linewidth far
from $\rm T_N$ is given approximately by
\begin{equation}
\Gamma_{q}({\rm HWHM}) = \frac{\pi {\rm R}_{0}^{\prime}
\rho^{*}}{4 \mu^{*}{\rm R}_{0}^{2}}\;q^*
\epsilon_{q}\,\frac{(1+\sigma)^2}{[1+\beta (1+\sigma)J(0){\rm
R}_{0}^{\prime}]},
\end{equation}
\noindent where $\epsilon_{q}$ is the magnon energy, $q^* = 2\pi
q/a$, $\mu^*$ = 2.969 \AA$^2$, $\rho^*$ = 5.864 \AA$^3$, and
$\beta$ = 1/k$_{\rm B}$T, with k$_{\rm B}$ the Boltzmann constant
\cite{stinch74}. The anisotropy parameter $\sigma$ is equal to
0.0184. The exchange parameter $J(0)$ = 6.02 meV includes both
first- and second-nearest neighbor exchange interactions.  The
parameters R$_0$ and R$_{0}^{\prime}$, which are both
temperature-dependent, can each be evaluated using either
experimental data or results from mean-field theory
\cite{stinch74}, leading to considerable differences in the
magnitude of the calculated linewidth and in its variation with
temperature. In determining R$_0$, we used experimental data for
the staggered magnetization \cite{jaccarino65}.  Calculation of
R$_0$ from the Brillouin function  produces linewidth values which
agree at the lowest temperatures and begin to deviate with
increasing temperature: at 40 K, the calculated linewidth is 11\%
smaller. For R$_{0}^{\prime}$, we used the derivative of the
Brillouin function. Calculation of R$_{0}^{\prime}$ instead from
experimental data for the parallel magnetic susceptibility
\cite{foner63} produces linewidth results which are 40\% larger at
15 K and 30\% smaller at 40 K.

Given the prominence of the magnon-magnon scattering channel in
the literature, the excellent agreement between this model
calculation and the experimental data is surprising. The dominance
of the relaxation by longitudinal fluctuations is, however,
consistent with arguments by Reinecke and Stinchcombe, who
estimated that the contribution to the linewidth from 4-magnon
scattering is only 1/$z$ of the magnitude of the above term
\cite{reinecke80}. Here, $z$ is the number of neighbors which
experience the strongest exchange interaction; $z$ = 8 for
MnF$_2$, for which case $z$ is the number of next-nearest
neighbors \cite{cscaveat}.
%In the calculation used
%here, the anisotropy is treated as anisotropic exchange in origin.
As the analytical expression on which the curves in Figs. 3 and 4
are based is valid only at low $q$, deviation from the data at
larger $q$ is not unexpected.  The general expression for the
linewidth resulting from scattering by longitudinal spin
fluctuations \cite{cs71b,stinch74} should be evaluated numerically
at high $q$ to see if the peak as a function of $q$ can be
reproduced \cite{cspredictedpeak,woolseypredictedpeak}.

An explanation of the peak centered at $q\sim 0.1$ r.l.u. (inset
in Fig. 3) requires a different mechanism \cite{curvaturecorr}. An
additional potential source of linewidth is the hyperfine
interaction, which gives rise to the scattering of electronic
magnons from nuclear spin fluctuations \cite{woolsey72}. The
contribution from the hyperfine interaction would only be weakly
temperature dependent, because the nuclear spin system is already
highly disordered thermally at 3 K. 4-magnon scattering terms in
which one electronic and one nuclear magnon interact have indeed
been shown to generate maxima in the linewidth at nonzero $q$, but
estimates of the amplitude of this contribution are at least an
order of magnitude smaller than the observed effect
\cite{woolsey72}.  The crossing of magnon and TA phonon modes at
$q\cong 0.04$ r.l.u. may also contribute to the
peak\cite{melcher70,smith77,montgomery73}. An additional
relaxation mechanism which must be considered as a possible source
of linewidth is that of magnon-phonon scattering. Experimental
estimates of the linewidth due to magnon-phonon relaxation in
MnF$_2$ in zero field range over three orders of magnitude, but
again appear too small to explain the observed peak
\cite{rotter90a,jongerden89,rotter90b}. A theoretical estimate of
the spin-lattice relaxation time (which should be of the same
order of magnitude as the magnon-phonon relaxation times)
corresponds to a linewidth of $\sim\!$ 0.5 $\mu$eV at 25 K in
MnF$_2$ \cite{cottam74b,wu82}.  In this theory, the magnon-phonon
interaction arises from the phonon modulation of the exchange
interaction, and is dominated by 2-magnon-1-phonon processes. The
result varies as T$^5$, which corresponds to a linewidth of 1 $
\mu$eV at 30 K and 5 $\mu$eV at 40 K. The maximum potential
contribution to our data would then be $\sim\!$ 60\% of the
linewidth at $q = 0$ and 40 K.  Other mechanisms which may
contribute to the presence of this peak include 2- and 3-magnon
non-momentum-conserving processes which originate from scattering
from defects \cite{loudon63,woolseyPhD,woolsey72}.  The linewidth
originating from the latter process is peaked at intermediate $q$.
Using parameters derived from comparison with data on RbMnF$_3$,
its contribution in MnF$_2$ can be estimated to be two orders of
magnitude smaller than the data \cite{woolseyPhD}.

%Using parameters derived from comparison with data on RbMnF$_3$,
%the latter contribution can be estimated to be no larger than
%0.007 $\mu$eV at 4 K in MnF$_2$, and should be peaked at
%intermediate $q$\cite{woolseyPhD}.

%Through the use of the new technique of neutron resonance
%spin-echo TAS, we have been able to measure low-temperature spin
%wave lifetimes over the full Brillouin zone with unprecedented
%accuracy.

%The excellent resolution of these linewidth measurements allows
%accurate comparison

The challenge to theory posed by the temperature- and
momentum-dependent peaks in the magnon linewidth in MnF$_2$ should
stimulate new activity in the field of spin wave decay mechanisms.
High-resolution lifetime measurements over the full Brillouin zone
in a relatively simple antiferromagnet such as MnF$_2$ permit
detailed evaluation of proposed processes, which should provide a
basis for addressing such interactions in more complex magnetic
systems.

% Your references go at the end of the main text, and before the
% figures.  For this document we've used BibTeX, the .bib file
% scibib.bib, and the .bst file Science.bst.  The package scicite.sty
% was included to format the reference numbers according to *Science*
% style.

\bibliography{scibib}

\begin{thebibliography}{10}
\bibitem{units}  We use a notation in which the momentum ${\bf Q}$ transferred by the
neutron is ${\bf Q} = {\bf q}+{\bf G}$, where ${\bf q}$ is the
momentum transfer within the Brillouin zone centered at the
reciprocal lattice vector ${\bf G}$. These quantities are
expressed in reciprocal lattice units (r.l.u.).  For instance,
${\bf Q}$=(HKL), with ${Q_a} = H (2\pi/a)$, and ${\bf q}=(hkl)$,
with ${q_c} = l (2\pi/c)$.
\bibitem{cscaveat} In contrast, in earlier work Cottam and Stinchcombe suggest that in the general case,
the linewidth contribution from 4-magnon scattering dominates at
low temperatures \cite{cs71b}.
\bibitem{curvaturecorr} The effect of the curvature of the magnon dispersion on the
spin-echo resolution is largest at $q = 0$, and largest at low
temperature, where the spin-wave stiffness is strongest.  Thus,
curvature not properly taken into account in the analytical
correction calculation would produce an apparent linewidth.
However, this spurious effect would be peaked at $q=0$, and is
therefore unlikely to be responsible for the observed peak.
\bibitem{cspredictedpeak}  A suggestion of non-monotonic behavior in the linewidth as a
function of $q$ is contained in Ref. \cite{cs71b}. For the case in
which the anisotropy takes the form of an anisotropy field $H_A$,
the linewidth peaks at an intermediate value of $q$ and then
declines as it approaches the Brillouin zone boundary at $q = 0.5$
r.l.u.  However, the estimated value of $q$ at which the peak occurs
would be only $\sim\!$ 0.008 r.l.u. for MnF$_2$. The behavior of the
linewidth at large $q$ for the case of anisotropic exchange was not
evaluated.
\bibitem{woolseypredictedpeak} Woolsey and White \cite{woolsey72} predict
a peak as a function of $q$ in MnF$_2$ for the linewidth due to a
4-magnon scattering process in a relatively small magnetic field
between 2 and 3 K. The peak becomes less pronounced and shifts to
higher $q$ as the temperature increases to 3 K (from 0.30 r.l.u.
at 2 K to 0.37 r.l.u. at 3 K). At a slightly higher field, only a
simple maximum is present between 4 and 10 K for a 4-magnon
scattering process \cite{woolseyPhD}. The authors did not
incorporate anisotropy in their calculations; later work by the
same authors showed that the inclusion of anisotropy results in a
decrease in the calculated linewidth of roughly an order of
magnitude for $q = 0$ in a large magnetic field \cite{white74a}.
\end{thebibliography}

\bibliographystyle{Science}

% Following is a new environment, {scilastnote}, that's defined in the
% preamble and that allows authors to add a reference at the end of the
% list that's not signaled in the text; such references are used in
% *Science* for acknowledgments of funding, help, etc.

%\begin{scilastnote}
%\item We would like to express our thanks to G. Schmidt of the Crystal Growth Facility of the
% Cornell Center for Materials Research for the loan of a MnF$_2$
%crystal of excellent quality, R. Henes and J. Major for the
%gamma-ray diffractometry measurements, J. Peters for cryogenic
%assistance, G. Khaliullin and R.K. Kremer for illuminating
%discussions, P. Aynajian for participation in some of the
%calibration measurements, and R. Noack for technical assistance.
%\end{scilastnote}

% For your review copy (i.e., the file you initially send in for
% evaluation), you can use the {figure} environment and the
% \includegraphics command to stream your figures into the text, placing
% all figures at the end.  For the final, revised manuscript for
% acceptance and production, however, PostScript or other graphics
% should not be streamed into your compliled file.  Instead, set
% captions as simple paragraphs (with a \noindent tag), setting them
% off from the rest of the text with a \clearpage as shown  below, and
% submit figures as separate files according to the Art Department's
% instructions.

\clearpage

%\noindent {\bf Fig. 1.} Please do not use figure environments to set
%up your figures in the final (post-peer-review) draft, do not include graphics in your
%source code, and do not cite figures in the text using \LaTeX\
%\verb+\ref+ commands.  Instead, simply refer to the figure numbers in
%the text per {\it Science\/} style, and include the list of captions at
%the end of the document, coded as ordinary paragraphs as shown in the
%\texttt{scifile.tex} template file.  Your actual figure files should
%be submitted separately.

\noindent{\bf Supporting Online Material} \newline
www.sciencemag.org
\newline Materials and Methods
\newline Fig. S1
\newline References

\parskip12pt
\subsection*{\bf Figures}
\begin{figure}
\includegraphics[width=1.00\linewidth]{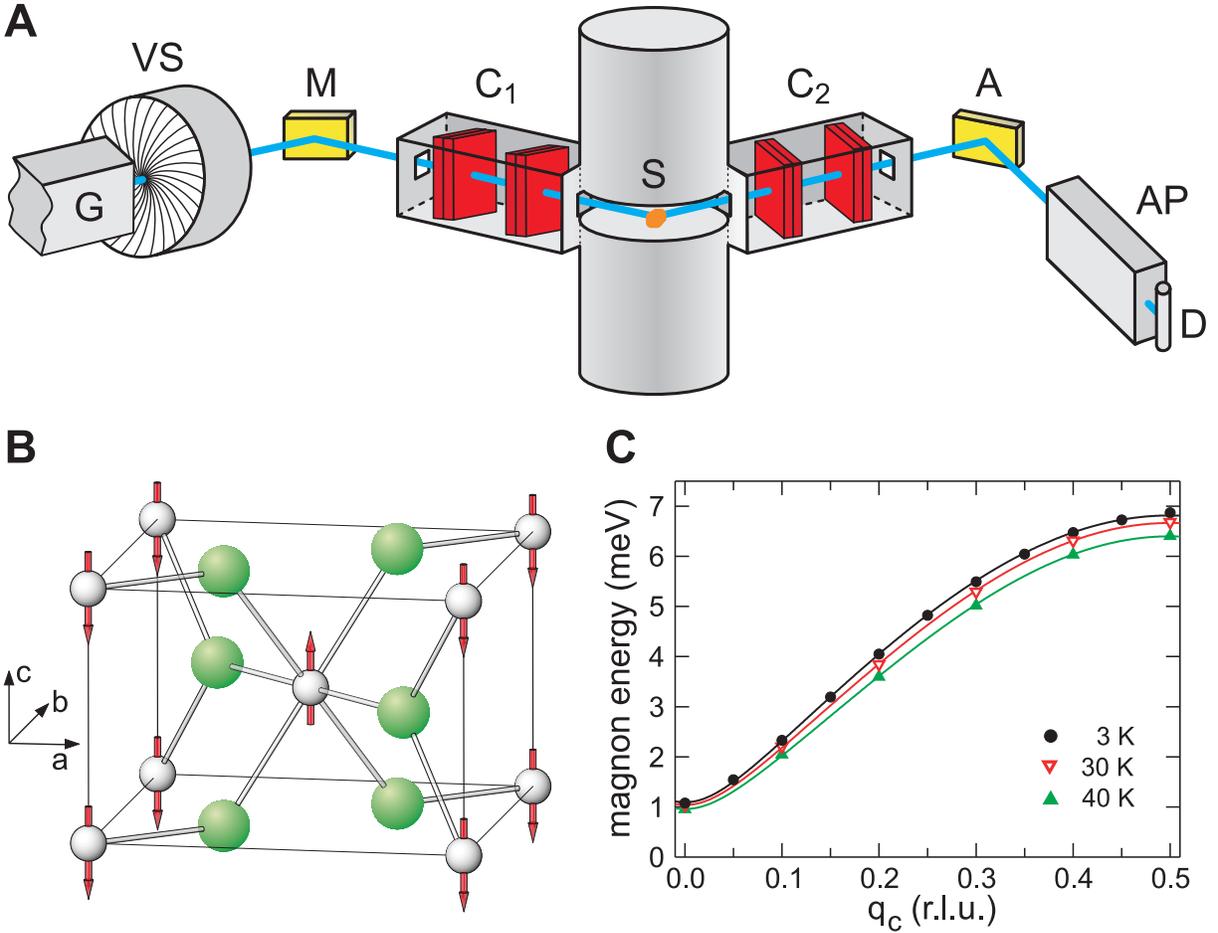}
\caption{{\bf (A)} A diagram of the spectrometer TRISP at the
FRM-II.  G denotes the polarizing guide and AP the transmission
polarizer; M and A are the monochromator and analyzer, as in TAS.
S is the sample and D the detector; VS indicates the velocity
selector. The resonance coil pairs (C$_1$ and C$_2$) are shown in
red, and the mu-metal shielding boxes which enclose them in gray.
The blue ray represents the path of the neutrons through the
spectrometer, from left to right on the diagram.  {\bf (B)} The
crystal and magnetic structure of MnF$_2$. The gray (smaller)
spheres represent Mn$^{2+}$ ions and the green (larger) spheres
the F$^-$ ions.  The arrows indicate the relative directions of
the Mn$^{2+}$ spins on the respective sublattices. {\bf (C)} The
magnon dispersion along the $q_c$ direction at three selected
temperatures at and below 40 K.  The data was taken on TRISP
during the course of the linewidth measurements. The curves show
the results of fits based on the same spin-wave result used by
Okazaki {\it et al.} \cite{okazaki64}, in which the anisotropy is
expressed by a single-ion form, and in which the interactions of
up to third-nearest neighbors are taken into account. }
\end{figure}

\begin{figure}
\includegraphics[width=1.00\linewidth]{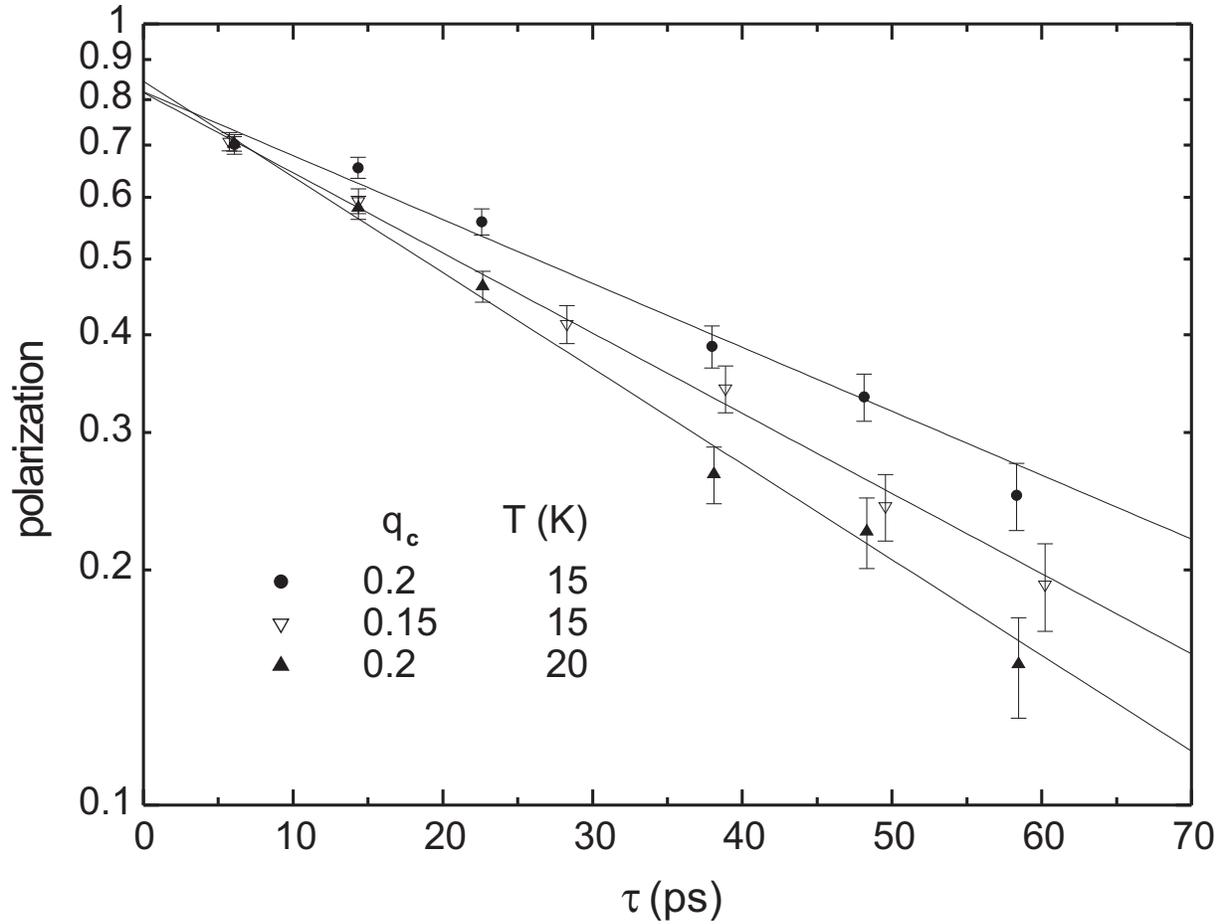}
\caption{Raw polarization data taken at (uppermost data) $q$ = 0.2
r.l.u. and T = 15 K, (middle) $q$ = 0.15 r.l.u. and T = 15 K, and
(lowest) $q$ = 0.2 r.l.u. and T = 20 K. The lines are exponential
fits to the data. The corresponding Lorentzian magnon linewidths
(HWHM) are $12.4\pm 0.8$ $\mu$eV, $15.6\pm 0.9$ $\mu$eV, and
$18.6\pm 1.0$ $\mu$eV, respectively. }
\end{figure}

\begin{figure}
\includegraphics[width=1.00\linewidth]{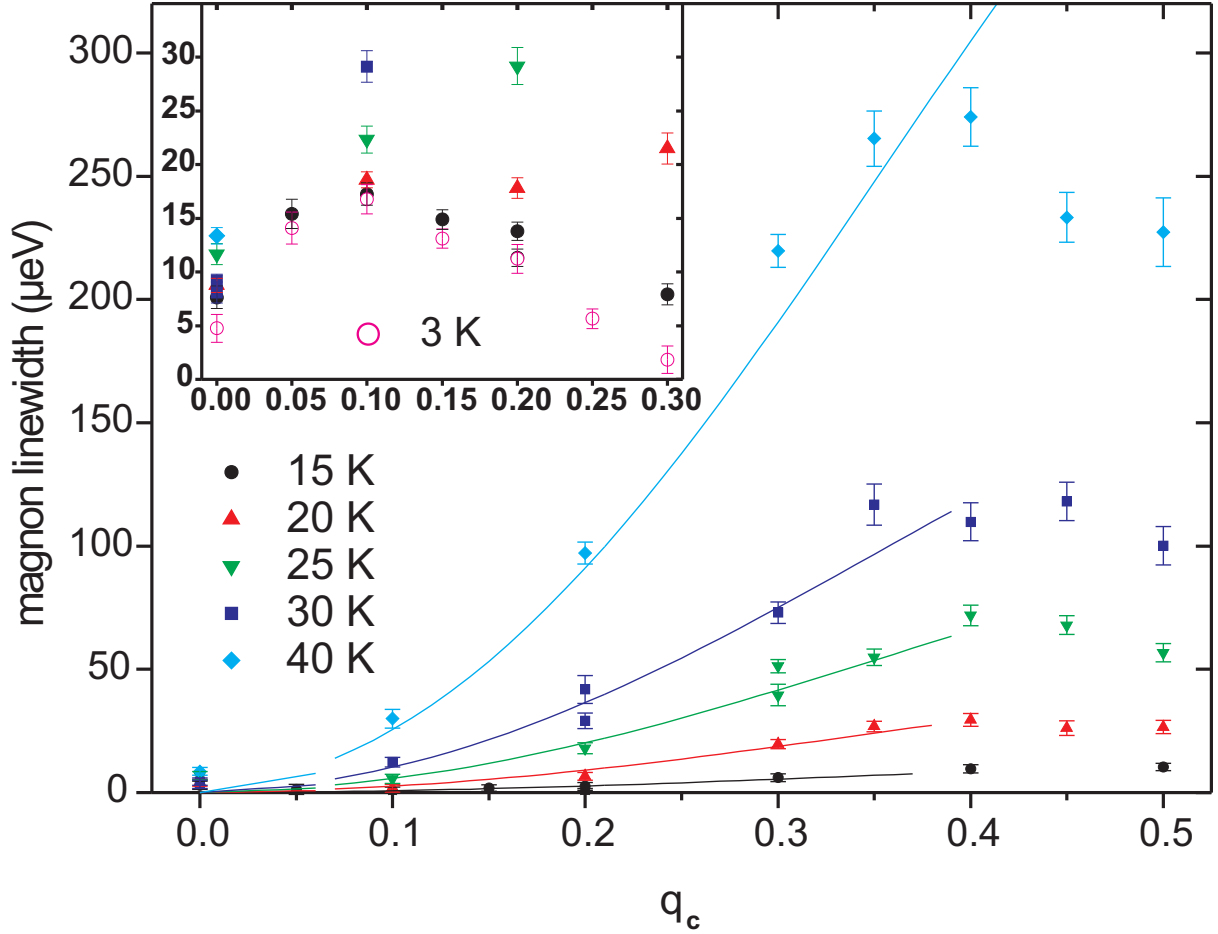}
\caption{Intrinsic magnon linewidth LW at temperatures ranging
from 15 - 40 K, as a function of $q$.  We have plotted (LW(T,
$q$)-LW(3 K, $q$)), where LW(3 K, $q$) is given in the inset (see
text).
%corrected data taken at 3 K and LW(T, $q$) is corrected
%data taken with T$>$ 3 K.
The curves show theoretical expressions from Refs. \cite{cs71b} and
\cite{stinch74} (see text).  Two different theoretical expressions
are valid for the small-$q$ case, depending on the magnitude of $q$
relative to the anisotropy energy.  Both expressions apply only to
small $q$; Stinchcombe and Reinecke have applied one of them to data
extending up to $q = 0.2$ r.l.u. for MnF$_2$ near $\rm T_N$
\cite{stinch74}. However, except at $q = 0$, the theory provides an
excellent fit to the data as a function of $q$ up to $q \sim 0.35$
r.l.u., both in the magnitude and in the $q$-dependence.
%Inset: corrected linewidth
%data plotted for small values of $q$ at temperatures ranging from
%3 - 40 K, without subtraction of the corrected 3 K data.  The axes
%correspond to those of the main panel.
}
\end{figure}

\begin{figure}
\includegraphics[width=1.00\linewidth]{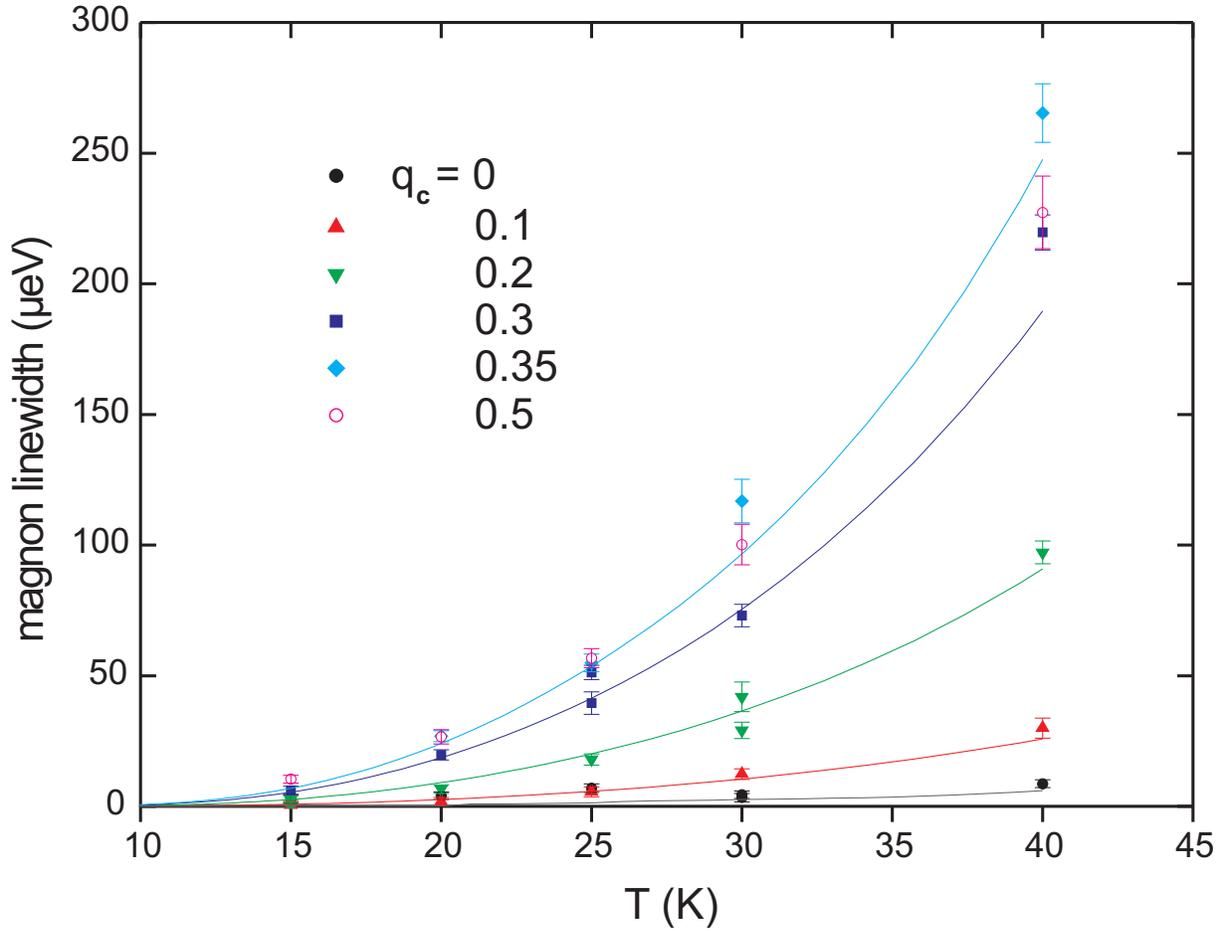}
\caption{Intrinsic magnon linewidth with $0 \leq q \leq 0.35$
r.l.u. and $q = 0.5$ r.l.u., shown as a function of temperature.
As in Fig. 3, data taken at 3 K has been subtracted. For $0.1 \leq
q \leq 0.35$ r.l.u., the curves show theoretical results from Ref.
\cite{stinch74}.
%The temperature dependence of the data is
%stronger at low $q$ than that predicted by this theory: the data
%is described reasonably well between 15 and 40 K by a power-law
%exponent of $3.8\pm 0.2$ for $q = 0.2$.  This exponent decreases
%to $3.3\pm 0.1$ for $q = 0.35$, which compares well with the
%theory . Even at values of $q$ larger than 0.35 r.l.u., the
%temperature dependence of the data is not far off from that
%predicted by the theory: the power-law exponent (which describes
%both theory and experiment adequately) differs by at most 0.1 from
%the approximate theoretical result of 3.3. However, for $q >
%0.35$, the magnitude of the predicted linewidth does not reflect
%the existence of the high-$q$ peak as a function of $q$
%\cite{cspredictedpeak,woolseypredictedpeak}.
For $q = 0$, results from Harris {\it et al.} are plotted
\cite{harris71}. }
\end{figure}

\end{document}